\magnification=1200
\baselineskip=18truept
\input epsf

\def\dirac{{\bf\rm D}}
\def\wilson{{\bf\rm B}}
\def\chiral{{\bf\rm C}}
\def\ham{{\bf\rm H}}
\def\mbham{{\cal H}}
\def\pfaff{{\bf\rm Pf}}
\def\det{{\bf\rm det}}


\def\draftversion{N}

\if \draftversion Y


\fi

\rightline{DPNU-97-25}
\rightline{May 1997}
\medskip

\vskip 2truecm
\centerline{\bf Lattice Formulation of Supersymmetric Yang-Mills Theories 
without Fine-Tuning}
\vskip 1truecm
\centerline{Nobuhito Maru${}^a$ and Jun Nishimura${}^b$}
\vskip 1truecm
\centerline {Department of Physics, Nagoya University}
\centerline {Chikusa-ku, Nagoya 464-01, Japan}
\vskip 1truecm
\centerline{${}^a$ e-mail: {\tt maru@eken.phys.nagoya-u.ac.jp}}
\centerline{${}^b$ e-mail: {\tt nisimura@eken.phys.nagoya-u.ac.jp}}
\vfill
\centerline{\bf Abstract}
\vskip 0.75truecm
We present a lattice formulation which gives
super Yang-Mills theories in any dimensions with simple supersymmetry
as well as extended supersymmetry 
in the continuum limit without fine-tuning.
We first formulate super Yang-Mills theories with simple supersymmetry
in 3,4,6 and 10 dimensions,
incorporating the gluino on the lattice using the overlap formalism.
In 4D, exact chiral symmetry forbids gluino mass, which ensures
that the continuum limit is supersymmetric without fine-tuning.
In 3D, exact parity invariance plays the same role.
6D and 10D theories being anomalous, we formulate them 
as anomalous chiral gauge theories as they are.
Dimensional reduction within lattice formalism 
is then applied to the theories in
3,4,6 and 10D to obtain super Yang-Mills theories in arbitrary dimensions 
with either simple or extended supersymmetry.

\bigskip
\leftline{PACS: 11.15.Ha; 11.30.Pb}
\leftline{Keywords: Lattice gauge theory, Supersymmetry}
\medskip

\vfill\eject

\centerline{\bf I. Introduction}
\medskip
Lattice formulation of supersymmetric gauge theories will provide a
numerical method to obtain non-perturbative results in these theories
and would complement the recent analytical developments [1,2]. Since
the lattice formulation breaks the continuous rotational and translational 
invariance, it is also expected to break supersymmetry. 
A practical strategy might be, therefore, to give up manifest supersymmetry 
on the lattice and to recover it in the continuum limit by fine-tuning,
as was advocated by Curci and Veneziano [3] some time ago.
They showed within lattice perturbation theory
that four-dimensional ${\cal N}=1$ super Yang-Mills theory
can be obtained by using the Wilson-Majorana fermion for the gluino
and fine-tuning the hopping parameter to the chiral limit.
Based on this work, some numerical simulations have been started [4].
An extension of Ref. [3] to ${\cal N}=2$ case has also been discussed [5].
Fine-tuning necessary in either case, 
however, might be a practical obstacle in extracting 
interesting non-perturbative features due to supersymmetry through numerical
simulations.

In general, thanks to the universality of the field theory,
one can hope to obtain a supersymmetric theory by fine-tuning
as many parameters as the relevant operators that breaks supersymmetry
around the supersymmetric ultraviolet fixed point.
If we have some symmetry that forbids those supersymmetry breaking operators,
we can avoid fine-tuning by imposing the symmetry on the lattice theory.
Indeed, this might be the best we can hope for concerning 
the supersymmetry on the lattice,
considering that even the continuous rotational and 
translational invariance can be restored only in the continuum limit, 
but without any fine-tuning, so long as we maintain the discrete rotational 
and translational invariance on the lattice.

In four-dimensional ${\cal N}=1$ super Yang-Mills theory,
the gluino is Majorana fermion in the adjoint representation of the gauge
group.
There are two relevant operators, which correspond to 
the gauge coupling and the gluino mass.
We can therefore obtain the supersymmetric continuum limit
by fine-tuning the gluino mass
to zero, as was shown in Ref. [3].
Alternatively, one can impose chiral symmetry, which forbids the
gluino mass, so that one can avoid fine-tuning.
Unfortunately the chiral symmetry is not easy to impose on the lattice.
There is even a no-go theorem under some modest assumptions [6].
We have, however, a formalism which preserves exact chiral symmetry;
that is the overlap formalism [7], which was originally developed 
to deal with chiral gauge theories on the lattice.
The key to circumvent the no-go theorem lies in the fact that the formalism
can be thought of as involving an infinite number of fermion fields at
each space-time point [8], which is beyond the assumptions of the no-go 
theorem.
In fact, in section 9 of [7], it has been suggested that
the overlap formalism can be used to formulate
four-dimensional ${\cal N}=1$ super Yang-Mills theory without fine-tuning.
A method in the same spirit using the domain-wall approach [9]
has been proposed in [10].

In this paper, we show that using the overlap formalism,
not only the four-dimensional ${\cal N}=1$ case, but also
any other super Yang-Mills theories in arbitrary dimensions 
with simple or extended supersymmetry can be accessible on the lattice
{\it without fine-tuning}.
We first consider super Yang-Mills theories 
in 3,4,6 and 10 dimensions [11], in which the dynamical degrees of freedom
of the gluon and the gluino exactly balance.
The gluino should be Majorana, Majorana (or equivalently Weyl), Weyl and
Majorana-Weyl, respectively.
As in Ref. [12], by means of dimensional reduction, one can obtain
all the other super Yang-Mills theories in any dimensions with
either simple or extended supersymmetry.
Although the parent theories in 6D and 10D suffer from gauge anomaly
[13] since the gluino is chiral,
the descendant theories are anomaly free
since the gluino, after dimensional reduction, is no longer chiral.
Our strategy is to formulate the parent theories 
in 3,4,6 and 10 dimensions using the overlap formalism
and to perform dimensional reduction on the lattice
to obtain all the descendant theories.

The overlap formalism, as a regularization of chiral gauge 
theories, gives the gauge anomaly correctly residing only in the phase 
of the fermion determinant.
The gauge dependence is expected to disappear in the continuum limit 
if and only if the fermion content is chosen to be anomaly free.
Indeed this seems to be natural enough to be true.
An obvious undesirable point of the formalism, however,
is that the gauge invariance of the phase for anomaly-free case
is not manifest on the lattice,
although we do not know at present whether it is possible at all
to define anomaly-free chiral gauge theories in a non-perturbative way
with manifest gauge invariance.
Apart from this, numerical simulation of 
chiral gauge theories is rather difficult in any case,
since the fermion determinant is complex in general.
When we apply the overlap formalism to the present case, however,
we are almost free from these problems, as we will see,
essentially because the theories we aim at are vector-like.

Let us start with three-dimensional ${\cal N}=1$ theory.
We should note here that the gluon can acquire mass without violating
gauge invariance in odd dimensions through Chern-Simons term.
Supersymmetry only requires the gluon and the gluino mass to be equal.
The massive case is referred to as
``supersymmetric Yang-Mills Chern-Simons theory'' 
in the literature [14].
In this paper, we consider only the massless case,
which can be obtained by imposing the parity invariance 
since it prohibits both the gluon and the gluino mass.
Parity invariance of three-dimensional gauge-fermion system has been
elucidated in Ref. [15], where it is emphasized that
the well-known parity anomaly is regularization dependent,
and that one can even preserve the parity invariance 
within gauge-invariant regularizations.
While the Wilson fermion breaks the parity invariance
because of the Wilson term and 
gives rise to parity anomaly in the massless limit [16],
the overlap formalism applied to three-dimensional massless Dirac fermion 
is shown to be parity invariant [15].
Here we extend the formalism to massless Majorana fermion and show that
it is parity invariant.
Using this formalism, we can obtain three-dimensional ${\cal N}=1$ 
super Yang-Mills theory in the continuum limit without fine-tuning.

Two-dimensional super Yang-Mills theory can be obtained through
dimensional reduction of the three-dimensional theory [17].
The theory now contains a scalar field in the adjoint representation,
which comes from the gauge field in the reduced direction.
This theory cannot be constructed directly in two dimensions,
since there are infinitely many relevant operators that break 
supersymmetry,
due to the existence of the scalar field which is dimensionless, and 
we do not have any symmetry on the lattice that can prohibit all of them.
We can, however, exploit the fact that it can be obtained 
through dimensional reduction of the three-dimensional theory. 
We discuss how this can be done within the lattice formalism.

The anomalous six- and ten-dimensional parent theories
can be formulated using the overlap formalism, 
in which only the phase of the fermion determinant is 
gauge dependent as it should.
After dimensional reduction, the gauge dependence is expected to
disappear, thus resulting in anomaly-free supersymmetric theories.
It is interesting to examine whether there exist non-trivial 
ultraviolet fixed points in more than four dimensions, 
thanks to supersymmetry.
Also of particular interest is 
that four-dimensional ${\cal N}=2$ and ${\cal N}=4$ 
super Yang-Mills theories, for example,
can be accessible in this way without fine-tuning.

This paper is organized as follows.
In section II, we consider
four-dimensional ${\cal N}=1$ super Yang-Mills theory.
We stress that, in this case, the fermion determinant 
obtained through the overlap formalism is real and
the gauge invariance on the lattice is guaranteed up to the sign.
In section III, we consider three-dimensional
${\cal N}=1$ super Yang-Mills theory.
We extend the overlap formalism constructed for
three-dimensional massless Dirac fermion to
massless Majorana fermion.
We prove that the formalism preserves the parity invariance.
We also show that the fermion determinant is real,
and the gauge invariance on the lattice 
is guaranteed up to the sign.
In section IV, we consider how we can obtain two-dimensional 
${\cal N}=1$ super Yang-Mills theory from the three-dimensional theory
constructed above, through dimensional reduction on the lattice.
In section V, we formulate the anomalous parent theories
in six and ten dimensions on the lattice, from which we obtain anomaly-free
super Yang-Mills theories through the lattice dimensional reduction.
Section VI is devoted to summary and future prospects.

\bigskip
\centerline{\bf II. 4D ${\cal N}=1$ Super Yang-Mills theory}
\medskip

Four-dimensional ${\cal N}=1$ super Yang-Mills theory can be obtained
using the overlap formalism without fine-tuning [7], 
where the gluino is treated as Weyl fermion.
Note that in four dimensions, in general, 
Weyl fermion in a real representation is equivalent to Majorana fermion.
Unlike in the Wilson's formalism applied to Majorana fermion, 
chiral symmetry is exact when the overlap formalism 
is used regarding the Majorana fermion as Weyl fermion.
Thus it provides a way to recover supersymmetry in the continuum limit 
without fine-tuning.

Here we note that when 4D Weyl fermion is in a real representation,
as in the present case, the fermion determinant should be real.
This is because 4D Weyl fermion in a real representation
is equivalent to Majorana fermion, 
whose fermion determinant should be the square root 
of that for the Dirac fermion, which is real positive.
In this section, we review that
the overlap formalism indeed gives a real value for the fermion determinant
of 4D Weyl fermion in a real representation.
The statement is commented at the footnote on page 315 of Ref. [7]
as a corollary of the lemma 4.5 in it.
We stress that this is important practically, since 
it means that the gauge invariance is 
guaranteed up to the overall sign even on the lattice.
The gauge dependence of the sign 
is expected to be very small on the lattice
and to disappear in the continuum limit.
Thus the overlap formalism gives 
almost gauge invariant regularization in this case.
This reflects the fact that the theory is actually vector-like.


In the overlap formalism [7], the determinant of 
a Weyl fermion is given by an
overlap of two many-body states. The two many-body states are
ground states of two many-body Hamiltonians describing non-interacting
fermions. Explicitly, the two many-body Hamiltonians on the lattice are
$$\mbham_\pm =a^\dagger \ham_\pm a,\eqno{(2.1)}$$
where
$$\ham_\pm = \pmatrix{ \wilson \pm m_o & \chiral \cr 
\chiral^\dagger & -\wilson \mp m_o \cr }, \eqno{(2.2)}$$
$$
\chiral(x\alpha i, y\beta j; U) = 
{1\over 2}
\sum_{\mu=1}^4 (\sigma_\mu)_{\alpha\beta} 
\Bigl [
\delta_{y,x+\hat\mu} (U_\mu(x))_{ij} - 
\delta_{x,y+\hat\mu} (U^\dagger_\mu(y))_{ij} 
\Bigr],\eqno{(2.3)}
$$
$$
\wilson(x\alpha i, y\beta j; U) = 
{1\over 2} \delta_{\alpha\beta}\sum_{\mu=1}^4  
\Bigl [
2 \delta_{xy}\delta_{ij} - 
\delta_{y,x+\hat\mu} (U_\mu(x))_{ij} - 
\delta_{x,y+\hat\mu} (U^\dagger_\mu(y))_{ij} 
\Bigr],
\eqno{(2.4)}
$$
and
$$
\sigma_1= \pmatrix{0 & 1 \cr 1 & 0 \cr},\ \ \ \
\sigma_2= \pmatrix{0 & -i \cr i & 0 \cr},\ \ \ \
\sigma_3= \pmatrix{1 & 0 \cr 0 & -1 \cr},\ \ \ \
\sigma_4=\pmatrix{i & 0 \cr 0 & i \cr}.\eqno{(2.5)}$$ 
$0< m_o < 1$ is a number that is kept fixed as the lattice spacing is
taken to zero. 
Note that $\wilson^\dagger = \wilson$.
We define $|\pm\rangle^{\rm WB}_U$ as the ground states of
the two Hamiltonians in (2.1) that obey the Wigner-Brillouin
phase choice [7], namely, 
${}^{\rm WB} _{\ \ 1}\!\langle
\pm|\pm\rangle^{\rm WB}_U$ are real and
positive for all $U$. 
Then the determinant of the Weyl fermion
is given by ${}^{\rm WB}_{\ \ U}\!\langle -|+\rangle^{\rm WB}_U$.
Note that the above condition on the phases of
$|\pm\rangle^{\rm WB}_U$ completely fixes the $U$ dependent part of 
them.\footnote{${}^\star$}
{Strictly speaking, the exceptions are the gauge configurations for which
${}^{\rm WB}_{\ \ 1}\!\langle \pm| \pm \rangle^{\rm WB}_U$ vanishes.
We can neglect them, however, since the measure of the set of those 
configurations is zero.}
Thus the fermion determinant is uniquely defined 
including its phase up to an irrelevant constant 
through the overlap formula for every gauge configuration.

In order to obtain the ground states of the many-body Hamiltonians,
we first diagonalize the single-particle Hamiltonians $\ham_\pm$
as
$$
\ham_\pm = V_\pm^\dagger \Lambda_\pm V_\pm ,
\eqno{(2.6)}
$$
where $\Lambda_\pm$ are real diagonal matrices 
and $V_\pm$ are unitary matrices.
Let us define the operators $a'_\pm$ through $a'_\pm=V_\pm a$.
Then the ground states of the many-body Hamiltonians can be obtained
by operating all the 
elements of ${a'_\pm}^\dagger$ that correspond to negative eigenvalues
on the kinematical vacuum $|0 \rangle$ annihilated by $a$.
We denote the states thus obtained as 
$|\pm\rangle_U$.
We define
$$
|\pm\rangle^{\rm WB}_U  = e^{i \chi_\pm (U)} |\pm\rangle_U,
\eqno{(2.7)}
$$
where $\chi_\pm (U)$ is chosen such that
${}^{\rm WB} _{\ \ 1}\!\langle
\pm|\pm\rangle^{\rm WB}_U$ are real and
positive for all $U$. 
The undetermined $U$ independent part of the phase is
irrelevant and can be fixed as one likes.


When the fermion is in a real representation, namely,
$U_\mu^\ast = U_\mu$,
we have
$$\Sigma \ham_\pm \Sigma^\dagger = \bigl[ \ham_\pm \bigr]^*,
\eqno{(2.8)}$$
where $\Sigma$ is a unitary matrix defined as
$$\Sigma= \pmatrix{\sigma_2 & 0 \cr 0 & -\sigma_2 \cr}.\eqno{(2.9)}$$
In deriving eq.(2.8),
we have used the property $-\sigma_2 \sigma_\mu \sigma_2 = \sigma_\mu^*$.
Using the identity, we can rewrite the many-body Hamiltonians as
$$\mbham_\pm =a^\dagger \ham_\pm a 
= b^\dagger \bigl[ \ham_\pm \bigr]^* b ,
\eqno{(2.10)}$$
where $b=\Sigma a$.
Note that
$\bigl[ \ham_\pm \bigr]^* $ can be diagonalized as follows.
$$
\bigl[ \ham_\pm \bigr]^*  = {{V_\pm}^*}^\dagger \Lambda_\pm {V_\pm}^* .
\eqno{(2.11)}
$$
Let us define the operators $b'_\pm$ through $b'_\pm=V^*_\pm b$
and construct the ground states of the many-body Hamiltonians
by operating all the components of
${b'_\pm}^\dagger$ that correspond to negative eigenvalues
on $|0 \rangle$ in the same order as we constructed 
$|\pm\rangle_U$.
We denote the states thus obtained as $\widetilde{|\pm\rangle}^{\rm }_U$. 
As before, we define
$$
\widetilde{| \pm\rangle}{}^{\rm WB}_U  = e^{i \tilde{\chi}_\pm (U)} 
\widetilde{|\pm\rangle}{}_U,
\eqno{(2.12)}
$$
where $\tilde{\chi}_\pm (U)$ is chosen such that
${}^{\rm WB}_{\ \ 1}\!\widetilde{\langle \pm } |
\widetilde{\pm\rangle}{}^{\rm WB}_U$
are real and positive for all $U$. 
Since the Wigner-Brillouin phase choice fixes the $U$ dependent part
of the phase of the ground states completely,
we can write as
$$
\widetilde{| \pm\rangle}{}^{\rm WB}_U  =
e^{i \theta_\pm} | \pm\rangle {}^{\rm WB}_U  ,
\eqno{(2.13)}
$$
where $\theta_\pm$ are constants independent of the gauge configuration
$U$.

As is obvious from the above construction, we have
$$
{}_1\!{}\langle \pm | \pm \rangle_U = 
\Bigl[ 
 {}_1\!{}\widetilde{\langle \pm} | \widetilde{\pm \rangle}{}_U 
\Bigr]^*
\eqno{(2.14)}
$$
$$
{}_U\langle - | + \rangle_U = 
\Bigl[ 
 {}_U\widetilde{\langle -} | \widetilde{+ \rangle}{}_U 
\Bigr]^*.
\eqno{(2.15)}
$$
Let us rewrite the above relations in terms of the ground states
with the Wigner-Brillouin phase choice.
>From eq. (2.14), we obtain 
$$
\eqalign{
e^{-i (\chi_\pm(U)-\chi_\pm(1))} \cdot
{}^{\rm WB}_{\ \ 1}\!\langle \pm | \pm \rangle^{\rm WB}_U 
& = \Bigl[
e^{-i (\tilde{\chi}_\pm(U)-\tilde{\chi}_\pm(1))} \cdot
 {}^{\rm WB}_{\ \ 1}\! \widetilde{\langle \pm } | \widetilde{\pm \rangle}
{}^{\rm WB}_U \Bigr]^* \cr
& =
e^{i (\tilde{\chi}_\pm(U)-\tilde{\chi}_\pm(1))}
 \Bigl[
 {}^{\rm WB}_{\ \ 1}\! \langle \pm |\pm \rangle {} ^{\rm WB}_U 
\Bigr]^*. \cr}
\eqno{(2.16)}
$$
Since 
${}^{\rm WB}_{\ \ 1}\!\langle \pm | \pm \rangle^{\rm WB}_U $
is real and positive, we have
$$
e^{-i (\chi_\pm(U)-\chi_\pm(1))}
= e^{i (\tilde{\chi}_\pm(U)-\tilde{\chi}_\pm(1))}.
\eqno{(2.17)}
$$
>From eq.(2.15), we have
$$
\eqalign{
e^{-i (\chi_{+}(U)-\chi_{-}(U))} \cdot
{}^{\rm WB}_{\ \ U}\!\langle - | + \rangle^{\rm WB}_U 
& = \Bigl[
e^{-i (\tilde{\chi}_{+}(U)-\tilde{\chi}_{-}(U))} \cdot
{}^{\rm WB}_{\ \ U}\! \widetilde{\langle - } | \widetilde{+ \rangle}
{}^{\rm WB}_U \Bigr]^* \cr
& =
e^{i (\tilde{\chi}_{+}(U)-\tilde{\chi}_{-}(U))}
e^{- i (\theta_{+} - \theta_{-} )}
 \Bigl[
 {}^{\rm WB}_{\ \ U}\! \langle - | + \rangle ^{\rm WB}_U 
\Bigr]^* , \cr}
\eqno{(2.18)}
$$
where we used eq.(2.13) at the second equality.
Using eq.(2.17), we obtain
$$
\eqalign{
{}^{\rm WB}_{\ \ U}\!\langle - | + \rangle^{\rm WB}_U 
& = 
e^{i (\chi_{+}(1)-\chi_{-}(1))}
e^{i (\tilde{\chi}_{+}(1)-\tilde{\chi}_{-}(1))}
e^{- i (\theta_{+} - \theta_{-} )} 
 \Bigl[
 {}^{\rm WB}_{\ \ U}\! \langle - | + \rangle ^{\rm WB}_U 
\Bigr]^* \cr
& = 
e^{i\theta} 
 \Bigl[
 {}^{\rm WB}_{\ \ U}\! \langle - | + \rangle ^{\rm WB}_U 
\Bigr]^*, \cr}
\eqno{(2.19)}
$$
where $\theta$ is a constant independent of the gauge configuration $U$.
Rewriting the above equation as
$$
e^{-i\theta/2} \cdot
{}^{\rm WB}_{\ \ U}\!\langle - | + \rangle^{\rm WB}_U 
 = 
 \Bigl[
e^{-i\theta/2} \cdot
 {}^{\rm WB}_{\ \ U}\! \langle - | + \rangle ^{\rm WB}_U 
\Bigr]^* ,
\eqno{(2.20)}
$$
we find that the overlap is real up to the irrelevant constant phase factor.

Thus we have proved that the fermion determinant defined 
by the overlap formula is real
when the 4D Weyl fermion is in a real representation,
which is indeed the case with the gluino in
4D ${\cal N}= 1$ super Yang-Mills theory.


\bigskip
\centerline{\bf III. 3D ${\cal N}=1$ Super Yang-Mills theory}
\medskip
 
In this section, we will see that 3D ${\cal N}=1$ Super Yang-Mills theory
can be obtained without fine-tuning using the overlap formalism.
The overlap formalism has been applied to odd dimensions 
for the first time in Ref. [15].
In it, it was shown that the overlap formalism gives a parity invariant 
lattice regularization of massless Dirac fermion in three dimensions.
Here we extend the formalism to massless Majorana fermion
and prove the parity invariance.
The parity invariance prohibits both gluon and gluino mass,
thus enabling a lattice formulation of 
3D ${\cal N}=1$ Super Yang-Mills theory without fine-tuning.
We also show that the fermion determinant is real as in the 4D case,
which is practically important since it guarantees the gauge invariance
up to the sign on the lattice.

The determinant of a single 3D massless Dirac fermion can be given by an
overlap of two many-body states [15].
The two many-body states are
ground states of two many-body Hamiltonians describing non-interacting
fermions, which can be given on the lattice as
$$\mbham_\pm =a^\dagger \ham_\pm a,\eqno{(3.1)}$$
where
$$\ham_\pm = \pmatrix{ \wilson \pm m_o & \dirac \cr 
-\dirac & -\wilson \mp m_o \cr }, \eqno{(3.2)}$$
$$
\dirac(x\alpha i, y\beta j; U) = 
{1\over 2}
\sum_{\mu=1}^3 (\sigma_\mu)_{\alpha\beta} 
\Bigl [
\delta_{y,x+\hat\mu} (U_\mu(x))_{ij} - 
\delta_{x,y+\hat\mu} (U^\dagger_\mu(y))_{ij} 
\Bigr],\eqno{(3.3)}
$$
$$
\wilson(x\alpha i, y\beta j; U) = 
{1\over 2} \delta_{\alpha\beta}\sum_{\mu=1}^3 
\Bigl [
2 \delta_{xy}\delta_{ij} - 
\delta_{y,x+\hat\mu} (U_\mu(x))_{ij} - 
\delta_{x,y+\hat\mu} (U^\dagger_\mu(y))_{ij} 
\Bigr].\eqno{(3.4)}
$$
$0< m_o < 1$ is a number that is kept fixed as the lattice spacing is
taken to zero. 
Note that $B^\dagger=B$ and $D^\dagger=-D$.

When the fermion is in a real representation, namely,
$U_\mu^\ast = U_\mu$,
we have
$$\Bigl[ \dirac \sigma_2 \Bigr]^t = -\dirac \sigma_2,\ \ \ \
\wilson^t=\wilson^*=\wilson.
\eqno{(3.5)}$$
We will see that in this case,
the many-body Hamiltonians split into two identical terms corresponding
to two Majorana fermions. The steps involved are similar to the ones in
[18].
We write out
$$a=\pmatrix{a_1\cr a_2\cr},
\eqno{(3.6)}$$
and define $\xi$, $\eta$ 
by the following Bogoliubov transformation.
$$\eqalign{
a_1 & ={1\over \sqrt{2}}\Bigl[\xi - i \eta\Bigr] \cr
a_2 & = {1\over \sqrt{2}}\sigma_2 \Bigl[ {\xi^\dagger}{}^t - 
i {\eta^\dagger}{}^t \Bigr]\cr}
\eqno{(3.7)}
$$
$\xi$ and $\eta$ obey canonical anticommutation relations.
Substitution of (3.7) in (3.1) and the use of
(3.5) results in 
$$\eqalign{
\pmatrix{a_1^\dagger & a_2^\dagger\cr}
\pmatrix{ \wilson \pm m_0 & \dirac \cr 
-\dirac & -\wilson \mp m_0 \cr } 
\pmatrix{a_1 \cr a_2\cr} & 
={1\over 2}\pmatrix{\xi^\dagger & \xi^t\cr}
\pmatrix{ \wilson \pm m_0 & \dirac \sigma_2\cr 
-\sigma_2 \dirac & -\wilson \mp m_0 \cr } 
\pmatrix{\xi \cr {\xi^\dagger}{}^t\cr} \cr
& + {1\over 2}\pmatrix{\eta^\dagger & \eta^t\cr}
\pmatrix{ \wilson \pm m_0 & \dirac \sigma_2\cr 
-\sigma_2 \dirac & -\wilson \mp m_0 \cr } 
\pmatrix{\eta \cr {\eta^\dagger}{}^t\cr}. \cr}\eqno{(3.8)}
$$
Both the many-body Hamiltonians on the righthand side of the above
equation are identical and each one of them stands for a single
Majorana fermion. The many-body Hamiltonians
for a single Majorana fermion are
$$\mbham_\pm^{(maj)} = 
{1\over 2} \pmatrix{\xi^\dagger & \xi^t\cr} 
\pmatrix{ \wilson \pm m_0 & \dirac \sigma_2\cr 
-\sigma_2 \dirac & -\wilson \mp m_0 \cr } 
\pmatrix{\xi \cr {\xi^\dagger}{}^t\cr}. 
\eqno{(3.9)}$$
We define $|\pm\rangle^{\rm WB}_U$ as the ground states of
the two Hamiltonians in (3.9) that obey the Wigner-Brillouin
phase choice, namely, 
${}^{\rm WB}_{\ \ 1}\!\langle \pm|\pm\rangle^{\rm WB}_U$ is real and
positive for all $U$. 
Then the determinant of a single Majorana fermion
is given by ${}^{\rm WB}_{\ \ U}\!\langle -|+\rangle^{\rm WB}_U$.
We comment that the decoupling of 3D massive Dirac fermion [15] 
in a real representation into two Majorana fermions can be
derived in a similar way.

Let us address the issue of parity invariance
in the context of the overlap
formalism of massless Majorana fermion [19].
Consider the parity transformed gauge field on the lattice, namely,
$$U^\prime_\mu(x) = U^\dagger_\mu (-x-\hat\mu) = U^t_\mu (-x-\hat\mu).
\eqno{(3.10)}$$
It follows from the definition of $\dirac$ and $\wilson$ in (3.3)
and (3.4) that 
$$\dirac(x\alpha i,y\beta j; U')=-\dirac(-x\alpha i,-y\beta j; U)
\eqno{(3.11)}$$
$$\wilson(x\alpha i,y\beta j; U')=\wilson
(-x\alpha i,-y\beta j; U).
\eqno{(3.12)}$$
If we now define 
$$\xi(x\alpha i) = i \eta(-x\alpha i)\eqno{(3.13)}$$
in (3.9), we have
$$\eqalign{ &
{1\over 2} \pmatrix{\xi^\dagger & \xi^t\cr} 
\pmatrix{ \wilson(U^\prime) \pm m_0 & \dirac(U^\prime) \sigma_2\cr 
-\sigma_2 \dirac(U^\prime) & -\wilson(U^\prime) \mp m_0 \cr } 
\pmatrix{\xi \cr {\xi^\dagger}{}^t\cr}
=\cr
&\ \ \ \ \ 
{1\over 2} \pmatrix{\eta^\dagger & \eta^t\cr} 
\pmatrix{ \wilson(U) \pm m_0 & \dirac(U) \sigma_2\cr 
-\sigma_2 \dirac(U) & -\wilson(U) \mp m_0 \cr } 
\pmatrix{\eta \cr {\eta^\dagger}{}^t\cr}. \cr}
\eqno{(3.14)}
$$

Since the many-body Hamiltonians for the gauge configuration $U$
and the ones for the parity transformed configuration $U'$ can be
written in the same form by an appropriate unitary transformation of the
fermion operators,
we can construct the many-body ground states for $U$ and $U'$ as
$$
 | {\pm \rangle}_U = f_{\pm}(\xi^\dagger)|0\rangle
\eqno{(3.15)}
$$
$$
 \widetilde{|\pm\rangle}_{U'} = f_{\pm}(\eta^\dagger)|0\rangle,
\eqno{(3.16)}
$$
where $|0\rangle$ is the kinematical vacuum annihilated by
$\xi$ (or equivalently by $\eta$) and 
$f_{\pm}(x)$ represent polynomials of $x$.
We have the following relations.
$$
{}_1\!{}{\langle \pm} | {\pm \rangle}{}_U = 
 {}_1\!{}\widetilde{\langle \pm} | \widetilde{\pm \rangle}{}_{U'}
\eqno{(3.17)}
$$
$$
{}_U\!{}\langle - | + \rangle{}_U = 
 {}_{U'}\!{}\widetilde{\langle -} | \widetilde{+ \rangle}{}_{U'}.
\eqno{(3.18)}
$$
Note that the gauge configuration $U=1$ remains the same under 
the parity transformation.

We define the ground states with the Wigner-Brillouin phase choice 
as follows.
$$
|\pm\rangle^{\rm WB}_U  = e^{i \chi_\pm (U)} |\pm\rangle_U,
\eqno{(3.19)}
$$
$$
\widetilde{|\pm\rangle}{}^{\rm WB}_U  = e^{i \tilde{\chi}_\pm (U)} 
\widetilde{|\pm\rangle}{}_U,
\eqno{(3.20)}
$$
where $\chi_\pm (U)$ and
$\tilde{\chi}_\pm (U)$ are chosen such that
${}^{\rm WB}_{\ \ 1}\!\langle \pm|\pm\rangle^{\rm WB}_U$
and ${}^{\rm WB}_{\ \ 1}\!\widetilde{\langle \pm}|
\widetilde{\pm\rangle}{}^{\rm WB}_U$
are real and positive for all $U$. 
As before, we can write as
$$
\widetilde{| \pm\rangle}{}^{\rm WB}_U  =
e^{i \theta_\pm} | \pm\rangle {}^{\rm WB}_U  ,
\eqno{(3.21)}
$$
where $\theta_\pm$ are constants independent of the gauge configuration
$U$.
>From eq.(3.17),
$$
\eqalign{
e^{-i (\chi_\pm(U)-\chi_\pm(1))}
{}^{\rm WB}_{\ \ 1}\!\langle \pm | \pm \rangle^{\rm WB}_U 
&= 
e^{-i (\tilde{\chi}_\pm(U')-\tilde{\chi}_\pm(1))}
 {}^{\rm WB}_{\ \ 1}\! \widetilde{\langle \pm }| \widetilde{\pm \rangle}
{}^{\rm WB}_{U'}  \cr
&= 
e^{-i (\tilde{\chi}_\pm(U')-\tilde{\chi}_\pm(1))}
 {}^{\rm WB}_{\ \ 1}\! \langle \pm | \pm \rangle
{}^{\rm WB}_{U'}.  \cr}
\eqno{(3.22)}
$$
Since 
${}^{\rm WB}_{\ \ 1}\!\langle \pm | \pm \rangle^{\rm WB}_U $
and
${}^{\rm WB}_{\ \ 1}\! \langle \pm | \pm \rangle ^{\rm WB}_{U'} $
are real positive, we have
$$
e^{-i (\chi_\pm(U)-\chi_\pm(1))}
= e^{-i (\tilde{\chi}_\pm(U')-\tilde{\chi}_\pm(1))}.
\eqno{(3.23)}
$$
>From eq.(3.18), we have
$$
\eqalign{
e^{-i (\chi_{+}(U)-\chi_{-}(U))} \cdot
{}^{\rm WB}_{\ \ U}\!\langle - | + \rangle^{\rm WB}_U 
& = 
e^{-i (\tilde{\chi}_{+}(U')-\tilde{\chi}_{-}(U'))} \cdot
{}^{\rm WB}_{\ \ U'}\! \widetilde{\langle - } |  \widetilde{+ \rangle}
{}^{\rm WB}_{U'} \cr
& =
e^{- i (\tilde{\chi}_{+}(U')-\tilde{\chi}_{-}(U'))}
e^{i (\theta_{+} - \theta_{-} )} \cdot
 {}^{\rm WB}_{\ \ U'}\! \langle - | + \rangle ^{\rm WB}_{U'}. \cr}
\eqno{(3.24)}
$$
Using eq.(3.23), we obtain
$$
\eqalign{
{}^{\rm WB}_{\ \ U}\!\langle - | + \rangle^{\rm WB}_U 
& = 
e^{i (\chi_{+}(1)-\chi_{-}(1))}
e^{- i (\tilde{\chi}_{+}(1)-\tilde{\chi}_{-}(1))}
e^{ i (\theta_{+} - \theta_{-} )} \cdot
{}^{\rm WB}_{\ \ U'}\! \langle - | + \rangle ^{\rm WB}_{U'} 
\cr
& = 
e^{i\theta} \cdot
{}^{\rm WB}_{\ \ U'}\! \langle - | + \rangle ^{\rm WB}_{U'} ,
\cr}
\eqno{(3.25)}
$$
where $\theta$ is a constant independent of the gauge configuration $U$.
By putting $U=1$, one finds $\theta=0$.
Therefore, we have
$$
{}^{\rm WB}_{\ \ U}\! \langle - | + \rangle {}^{\rm WB}_U = 
 {}^{\rm WB}_{\ \ U'}\! \langle - | + \rangle {}^{\rm WB}_{U'}.
\eqno{(3.26)}
$$

Thus the overlap formalism for 3D massless Majorana fermion
preserves the parity invariance.
We can therefore use it to 
obtain 3D ${\cal N}=1$ super Yang-Mills theory 
in the continuum limit without fine-tuning.


Let us next show that the fermion determinant is real. 
We first comment that this is formally expected in the continuum.
The fermion determinant of concern is actually the Pfaffian 
(The word ``fermion {\it determinant}'' might, therefore, be confusing 
in this sense, but we keep on using this term.) 
of the antisymmetric operator $\dirac \sigma_2$.
We note that
$$
(\dirac\sigma_2)^\ast = \sigma_2 (\dirac \sigma_2) \sigma_2.
\eqno{(3.27)}
$$
Taking the Pfaffian on both sides, we obtain
$$
\eqalign{
\bigl[\pfaff(\dirac \sigma)\bigr]^\ast  
& = \pfaff( \sigma_2 (\dirac \sigma_2) \sigma_2) \cr
& = - \det (\sigma_2) \pfaff(\dirac \sigma_2)  \cr
& =  \pfaff(\dirac \sigma_2),  \cr}
\eqno{(3.28)}
$$
which means that the Pfaffian is real.
>From the first line to the second line, we used the formula
$$
\pfaff(X^t A X) = \pfaff(A) \det(X),
\eqno{(3.29)}
$$
where $A$ is an antisymmetric matrix and $X$ is an arbitrary matrix.
We will see that the fermion determinant defined through
the overlap formula is indeed real
\footnote{$\! {}^\star$}
{We thank R.Narayanan for pointing this out to the authors.}.

Although the proof goes in a similar way as in the previous section,
things are a little more complicated due to the fact that the many-body 
Hamiltonians (3.9) for the Majorana fermion violate fermion number
and we have to make Bogoliubov transformation instead of simple unitary
transformation in order to obtain the ground states.
Let us write the many-body Hamiltonians for a single Majorana fermion as
$$\mbham_\pm^{(maj)} = 
{1\over 2} \pmatrix{\xi^\dagger & \xi^t\cr} \ham_\pm 
\pmatrix{\xi \cr {\xi^\dagger}{}^t \cr} .
\eqno{(3.30)}$$
Due to the hermiticity of $\mbham_\pm^{(maj)}$ 
and the anticommutation relations among the $\xi$ operators,
the hermite matrix $\ham_\pm$ must have the following particular form.
$$\ham_\pm = \pmatrix{h & \lambda \cr - \lambda^* & -h^*}, 
\eqno{(3.31)}$$
where $h$ is hermitian and $\lambda$ is antisymmetric.
This is indeed satisfied with the explicit form of $\ham_\pm$
given through (3.9).
Suppose $(x,y)^t$ is an eigenvector of $\ham_\pm$ with the eigenvalue
$\omega$, then one can easily see that 
$(y^*,x^*)^t$ is an eigenvector of $\ham_\pm$ with the eigenvalue
$-\omega$, due to (3.31).
One can therefore diagonalize the $\ham_\pm$ as follows.
$$
\ham_\pm = {V_\pm} ^\dagger \Lambda_\pm V_\pm 
\eqno{(3.32)}
$$
$$V_\pm =\pmatrix{x^\dagger  & y^\dagger  \cr
\vdots & \vdots \cr
 y^t  & x^t  \cr
\vdots & \vdots \cr}, 
\eqno{(3.33)}$$
where $\Lambda_\pm$=diag($\omega$,$\cdots$,-$\omega$,$\cdots$).
Without loss of generality, we can take the first half of the
diagonal elements to be positive.
Due to the particular form of the unitary matrix $V_\pm$,
the $\xi'_\pm$ operators, which satisfy canonical anticommutation
relations, can be consistently defined through
$$\pmatrix{{\xi'_\pm} \cr {{{\xi ' _\pm}}}{}^{\dagger t}  \cr} 
=V_\pm \pmatrix{\xi \cr {\xi ^\dagger}{}^t   \cr} .
\eqno{(3.34)}$$
This gives the desired Bogoliubov transformation.
The ground states 
of the many-body Hamiltonians
can be given by the states 
$|\pm\rangle_U$
annihilated by $\xi'_\pm$.
Note that the above states are different from the kinematical vacuum
$|0\rangle$ annihilated by $\xi$, since
${\xi'_\pm}$ and $\xi$ are connected 
through the Bogoliubov transformation 
instead of a simple unitary transformation.


In order to show that the fermion determinant is real, we
start with the following identity.
$$\Gamma \ham_{\pm} \Gamma ^\dagger= [\ham_\pm]^*,
\eqno{(3.35)}
$$
where $\Gamma$ is a unitary matrix defined by
$$\Gamma = \pmatrix{i \sigma_2 & 0 \cr 0 & i \sigma_2 \cr}. 
\eqno{(3.36)}
$$
By defining the $\eta$ operators through $\eta = i\sigma_2 \xi$,
one can rewrite the many-body Hamiltonian as
$$\mbham_\pm^{(maj)} = 
{1\over 2} \pmatrix{\xi^\dagger & \xi^t\cr} \ham_\pm 
\pmatrix{\xi \cr {\xi^\dagger}{}^t\cr} = 
{1\over 2} \pmatrix{\eta^\dagger & \eta^t\cr} [\ham_\pm]^* 
\pmatrix{\eta \cr {\eta^\dagger}{}^t\cr},
\eqno{(3.37)}$$
which can be diagonalized in terms of 
$$\pmatrix{\eta'_\pm \cr {{\eta'_\pm}}{}^{\dagger t}\cr} 
=V_\pm^* \pmatrix{\eta \cr {\eta^\dagger}{}^t\cr} .
\eqno{(3.38)}
$$
The ground states can now be given by
the states $\widetilde{|\pm\rangle}$ 
annihilated by ${\eta'_\pm}$. 
Due to eqs. (3.34) and (3.38), 
one can take the following particular forms for
the many-body ground states.
$$
|\pm \rangle_U = f_{\pm}(\xi^\dagger)|0\rangle
\eqno{(3.39)}
$$
$$
 \widetilde{|\pm\rangle}_{U} = f^*_{\pm}(\eta^\dagger)|0\rangle,
\eqno{(3.40)}
$$
where $f_{\pm}(x)$ represent polynomials of $x$ and 
$f_{\pm}^*(x)$ represent polynomials with the coefficients which
are complex conjugate of those of $f_{\pm}(x)$.
>From this, we have the following relations.
$$
{}_1\!{}{\langle \pm} | {\pm \rangle}{}_U 
= 
\Bigl[ 
 {}_1\!{}\widetilde{\langle \pm }| \widetilde{\pm \rangle}{}_{U}
\Bigr] ^* 
\eqno{(3.41)}
$$
$$
{}_U \langle - | + \rangle{}_U 
= 
\Bigl[ 
 {}_U \widetilde{\langle -} | \widetilde{+ \rangle}{}_{U}
\Bigr] ^* .
\eqno{(3.42)}
$$
The rest of the proof goes exactly in the same way
as in the previous section.

Thus we find that the overlap for the 3D massless Majorana fermion is real.
The gauge invariance on the lattice is therefore guaranteed 
up to the sign of the fermion determinant.
Also, as a corollary, we find that 
the overlap formula for massless Dirac fermion in a real representation
is real positive, since it is the square of that for massless Majorana 
fermion.

\bigskip
\centerline{\bf IV. 2D Super Yang-Mills theory through the lattice 
dimensional reduction}
\medskip

Super Yang-Mills theory in two dimensions can be obtained 
by dimensional reduction of the three-dimensional super Yang-Mills 
theory [17].
The gauge field in the reduced dimension becomes a scalar field
in two dimensions.
The key point of obtaining the supersymmetric continuum limit 
in three and four dimensions is that
there exists a symmetry that prohibits the supersymmetry breaking 
relevant operators.
Now in two dimensions, we have a scalar field and due to this,
there are infinitely many relevant operators that break supersymmetry.
We can, however, exploit the fact that two-dimensional super Yang-Mills
theory can be obtained by dimensional reduction of the 
three-dimensional super Yang-Mills theory.
The main issue here is how to perform dimensional reduction
on the lattice.

Dimensional reduction has been discussed intensively
in the context of finite temperature field theory [20,21].
Finite temperature field theory can be formulated by keeping 
the physical extent in one direction finite,
which corresponds to the inverse temperature,
while sending those in the other directions to infinity corresponding
to the infinite volume limit.
In the zero temperature limit, one obtains the original field theory
in the infinite volume naturally.
In the high temperature limit, after integrating out the oscillating modes
in the inverse temperature direction, one obtains an effective field theory
with one dimension lower than the original theory.
The effective field theory contains only local interactions [20].

The boundary condition in the inverse temperature direction is relevant 
to physics, since the physical extension in this direction is kept finite
in contrast to the other directions where the boundary condition does not
affect the physics due to the infinite volume limit.
In finite temperature field theory, 
the boundary condition in the inverse temperature direction should be
periodic for bosonic fields and anti-periodic for fermionic fields.
The dimensionally reduced theory obtained
in the high temperature limit is composed of bosonic fields only [20],
since fermionic fields do not have zero modes in the inverse 
temperature direction due to the boundary condition.
When we consider the dimensional reduction in the context of 
supersymmetric theories,
we have to take periodic boundary condition also for the fermionic fields
in order to preserve supersymmetry.
Hence, fermionic fields as well as bosonic fields appear in 
the resulting dimensionally reduced theory,
which should be a local theory with supersymmetry.

Dimensional reduction on the lattice can be done for supersymmetric
theories as well as for high-temperature limit of ordinary field theories 
[21].
We consider the dimensional reduction of three-dimensional theory
down to two-dimensional theory. 
In order to achieve the dimensional reduction, we have to make
the physical extent in one direction, say $l_3$, finite,
while we take 
the physical extent in the other directions, say $l$, to infinity,
corresponding to the infinite volume limit.
Let us denote the typical scale of the theory ({\it e.g.;} 
the inverse of the lambda parameter) as $r$.
In order to realize the dimensional reduction, we have to take
$l_3 \ll r \ll l$.

In addition to this, we have to take the continuum limit $a\rightarrow 0$,
since we are working on the lattice.
If we take the lattice size to be $L\times L \times L_3$,
we have $l=aL$ and $l_3=a L_3$.
A typical correlation length $\xi$ is related to $r$ through
$r=a \xi$.
The dimensional reduction on the lattice 
can therefore be realized
by taking $1 \ll L_3  \ll \xi  \ll L$.

In this way, 
one can obtain 2D ${\cal N}=1$ super Yang-Mills theory from
3D ${\cal N}=1$ super Yang-Mills theory in Section III.
The point in avoiding fine-tuning which seemed to be inevitable
when we consider 2D theory directly is that
we have the 3D rotational and translational invariance 
restored in the continuum limit at the scale much smaller
than the extent of the reduced direction.
Similarly,
one can dimensionally reduce the 4D ${\cal N}=1$ super Yang-Mills theory
in Section II, to obtain
3D ${\cal N}=2$ and 2D ${\cal N}=2$ super Yang-Mills theories.

\bigskip
\centerline{\bf V. Anomalous 6D and 10D theories and their
dimensional reductions}
\medskip

In this section, 
we discuss the application of 
the overlap formalism to anomalous 6D and 10D theories,
>from which we obtain anomaly-free super Yang-Mills theories
by the lattice dimensional reduction described in the previous section.
Before that, let us briefly review
how we get the dimensions 3,4,6 and 10.

Super Yang-Mills theory requires the balance between the bosonic
and fermionic degrees of freedom.
In $D$ dimensions the physical degrees of freedom of 
the gauge field is
$(D-2)$, while the physical degrees of freedom of a Dirac 
fermion is $2^{[D/2]}$, where $[x]$ denotes the largest integer 
not more than $x$.

In the following, we summarize
the relevant properties of the spinors in Minkowski space
with arbitrary dimensions.
In even dimensions, Dirac fermion is decomposed into two Weyl fermions.
In 2, 3 and 4 dimensions, Dirac fermion in a real representation 
decomposes into two Majorana fermions.
In 8 and 9 dimensions, massless Dirac fermion in a real representation
decomposes into two pseudo Majorana fermions.
In 4 dimensions, Weyl fermion in a real representation is equivalent
to Majorana fermion.
In 8 dimensions, Weyl fermion in a real representation is equivalent
to pseudo Majorana fermion.
In 2 dimensions, Weyl fermion in a real representation further
decomposes into two Majorana-Weyl fermions.
All the above statements hold for the dimensions up to modulo 8.

It is now easy to find that the dimensions in which the physical degrees
of freedom of the gauge field and the gluino balance are
3, 4, 6 and 10.
The gluino in each dimension is,
Majorana in 3D, Majorana (or equivalently Weyl) in 4D,
Weyl in 6D, Majorana-Weyl in 10D.
Note that in 6 and 10 dimensions the gluino is chiral, giving 
rise to gauge anomaly, which has been studied in Ref. [13].
Thus in these dimensions, super Yang-Mills theory cannot be considered as 
a consistent quantum field theory with unitarity.
It is not even supersymmetric actually, since the gauge mode
does not decouple, which adds to the bosonic degrees of freedom.
Still we can define it as a statistical system
by naively integrating out the gauge mode.
This theory can be formulated using the overlap formalism.
The overlap formalism for Majorana-Weyl fermion,
which is necessary in formulating the 10D theory,
is studied in Ref. [18], where its application to 10D super Yang-Mills
theory as a regularization of 4D ${\cal N}=4$ super Yang-Mills theory
is suggested.
Including this as a special case,
we give a general prescription to obtain anomaly-free super Yang-Mills
theories by dimensional reduction 
>from the anomalous parent theories in 6D and 10D.

We apply the lattice dimensional reduction considered in the 
previous section to the 6D and 10D theories.
After reducing one direction, we obtain 5D and 9D theories
with the gluino being Dirac and pseudo Majorana respectively.
Since now the fermions are no longer chiral, 
we do not have gauge anomaly.
This means that although 
the chiral determinant calculated through the overlap
formalism in 6D and 10D is generally complex 
and the phase is gauge dependent,
the gauge dependence disappears after the dimensional reduction
by taking the continuum limit.
Moreover, we know in the continuum that the fermion determinant for
the above theories in 5D and 9D should be real, 
which can be used as useful information in numerical simulations.
Note also that since the gauge mode decouples, the resulting 
dimensionally reduced theories must be supersymmetric.

Of course, it remains to be seen whether there exists a non-trivial 
ultra-violet fixed point which allows a continuum limit in more than
four dimensions, since gauge theory in more than four dimensions is 
perturbatively unrenormalizable.
This is an issue that has to be explored nonperturbatively. 
In non-supersymmetric Yang-Mills theory, numerical simulations
in more than four dimensions have given negative conclusions [22,23].
However, we might be able to have non-trivial ultra-violet fixed points
in supersymmetric theories, thanks to supersymmetry.
This is an interesting issue also in the context of 
string theory [23,24].

Apart from this, we can further dimensionally reduce the theories down
to four dimensions. From six dimensions, we obtain 
four-dimensional ${\cal N}=2$ super Yang-Mills theory, while from 
ten dimensions, we obtain four-dimensional ${\cal N}=4$ 
super Yang-Mills theory.
These theories should exist from perturbative point of view,
and therefore it is no doubt that we can obtain a continuum limit, 
in contrast to the theories in more than four dimensions.
We can further dimensionally reduce the theories even lower.

\bigskip
\centerline{\bf VI. Summary and future prospects}
\medskip

In this paper, we presented a method to deal with
supersymmetric Yang-Mills theory on the lattice
in any dimensions with either simple or extended supersymmetry 
without fine-tuning.
Instead of preserving supersymmetry manifestly on the lattice,
we impose other symmetries on the lattice that ensure the continuum limit
to be supersymmetric automatically.
This is quite analogous to how we deal with 
continuous translational and rotational symmetries on the lattice.
Although we break them on the lattice, we can restore them in 
the continuum limit without fine-tuning, so long as we maintain the discrete
translational and rotational symmetry on the lattice. 
As for supersymmetry, we have to prohibit the gluon and the gluino mass.
The former is prohibited in even dimensions by the gauge invariance.
In four dimensions, the chiral symmetry further prohibits the gluino mass,
while in six and ten dimensions, it is prohibited by 
the chiral nature of the gluino.
In three dimensions, we can prohibit both the gluon and the gluino mass
by imposing the parity invariance.
Note that all the necessary ingredients to restore supersymmetry 
automatically, namely the chiral symmetry in 4D, the parity invariance in 3D,
and the chiral nature in 6D and 10D, are 
what the standard lattice formalism of fermions fails to deal with.
Surprisingly enough, 
the overlap formalism deals with all of these features quite nicely, thus
enabling a lattice formulation of 
super Yang-Mills theories without fine-tuning.
We also argued that the dimensional reduction technique
within the lattice formalism, 
which has been developed in the context of finite-temperature
field theory,
can be used to obtain 
all the other 
super Yang-Mills theory in arbitrary dimensions with either single
or extended supersymmetry.
Although the 6D and 10D theories are anomalous and not supersymmetric
actually, the theories after dimensional reduction should
be both anomaly-free and supersymmetric.
We stressed that all the super Yang-Mills theories
in the above sense are vector-like and the fermion determinant should be real.
Hence we are almost free from the subtlety as with general anomaly-free
chiral gauge theories,
in which the phase can take arbitrary values and its gauge dependence 
within the overlap formalism is not restricted at all kinematically.

Let us comment on some possible applications of our formalism.
In four-dimensional ${\cal N}=1$ super Yang-Mills theory,
the gluino condensation is an important issue in phenomenology.
The argument of Witten index [25] and the instanton calculations [26]
suggest that the condensation indeed occurs [2].
It would be interesting to examine this in numerical simulations.
In four-dimensional ${\cal N}=2$ super Yang-Mills theory,
the scalar fields will have undetermined VEV's, which give rise
to a non-trivial moduli space.
We will have to fix the VEV's by hand in numerical simulations.
It would be interesting to examine the conjectures given in Ref. [1]
numerically.
Also an issue which deserves further study 
is the possible existence of non-trivial ultra-violet fixed points
in more than four dimensions in the supersymmetric case.
Considering the variety of the applications,
an efficient algorithm for the overlap formalism
is highly desired.

Finally we should comment that we cannot put additional matters 
in a supersymmetric way into the present formalism.
Thus the interesting conjectures on the infrared fixed points
in supersymmetric QCD [2] cannot be examined within the formalism
as it stands.
This might be possible, if we could extract some
remnant of the supersymmetry in the overlap formalism,
which is of course an interesting issue itself worth studying in future.

\bigskip
\bigskip
\centerline{\bf Acknowledgement}
\medskip

The authors would like to thank R. Narayanan for helpful communication
throughout this work.
J.N. is also grateful to T. Eguchi, N. Haba and P. Pouliot 
for valuable comments.


\vfill\eject
\centerline{\bf References}

\item{1.} 
N. Seiberg and E. Witten, Nucl. Phys. B426 (1994) 19; B431 (1994) 484. 
\item{2.} N. Seiberg, Nucl. Phys. B435 (1995) 129; K. Intriligator and 
N. Seiberg, Nucl. Phys. Proc. Suppl. 45BC (1996) 1.
\item{3.} G. Curci and G. Veneziano, Nucl. Phys. B292 (1987) 555.
\item{4.} I. Montvay, Nucl. Phys. B466 (1996) 259;
A. Donini and M. Guagnelli, Phys. Lett. B383 (1996) 301.
\item{5.} I. Montvay, Phys. Lett. B344 (1995) 176;
I. Montvay, Nucl. Phys. B445 (1995) 399.
\item{6.} H.B. Nielsen and M. Ninomiya, Nucl. Phys. B185 (1981) 20;
Nucl. Phys. B193 (1981) 173; B195 (1982) 541(E).
\item{7.} R. Narayanan and H. Neuberger, Nucl. Phys. B443 (1995) 305.
\item{8.} R. Narayanan and H. Neuberger, Phys. Lett. B302 (1993) 62;
Nucl. Phys. B412 (1994) 574. 
\item{9.} V. Furman and Y. Shamir, Nucl. Phys. B439, 54 (1995),
T. Blum and A. Soni, preprint hep-lat/9611030 (1996).
\item{10.} J. Nishimura, preprint, hep-lat/9701013, 
to appear in Phys. Lett. B.
\item{11.} See, for example, Appendix 4.A in 
{\sl Superstring Theory, vol. 1},
M.B. Green, J.H. Schwarz and E. Witten, 
Cambridge University Press (1987).
\item{12.} L. Brink, J.H. Schwarz and J. Scherk, Nucl. Phys. B121 (1977) 77.
\item{13.} 
P.H. Frampton and T.W. Kephart, Phys. Rev. Lett. 50 (1983) 1343,1347;
P.K. Townsend and G. Sierra, Nucl. Phys. B222 (1983) 493;
B. Zumino and Y.-S. Wu and A. Zee, Nucl. Phys. B239 (1984) 477.
\item{14.} See, for example, F. Ruiz Ruiz and P. van Nieuwenhuizen, 
Nucl. Phys. B486 (1997) 443.
\item{15.} R. Narayanan and J. Nishimura, preprint, hep-th/9703109.
\item{16.} A. Coste and M. L\"uscher, Nucl. Phys. B323 (1989) 631.
\item{17.} S. Ferrara, Lett. Nuovo Cim. 13 (1975) 629.
\item{18.} P. Huet, R. Narayanan and H. Neuberger, 
Phys. Lett. B380 (1996) 291.
\item{19.} R. Narayanan, private communication.
\item{20.} T. Reisz, Z. Phys. C53 (1992) 169.
\item{21.} 
P. Lacock, D.E. Miller and T. Reisz,
Nucl. Phys. B369 (1992) 501;
L. K\"arkk\"ainen, P. Lacock, D.E. Miller, B. Petersson and T. Reisz,
Phys. Lett. B282 (1992) 121;
L. K\"arkk\"ainen, P. Lacock, B. Petersson and T. Reisz,
Nucl. Phys. B395 (1993) 733.
\item{22.} H. Kawai, M. Nio, and Y. Okamoto,
Prog. Theor. Phys. 88, (1992) 341.
\item{23.} J. Nishimura, Mod. Phys. Lett. A11 (1996) 3049.
\item{24.} 
T. Banks, W. Fischler, S. Shenker and L. Susskind, preprint,
hep-th/9610043 (1996); 
N. Ishibashi, H. Kawai, Y. Kitazawa and A. Tsuchiya,
preprint, hep-th/9612115 (1996).
\item{25.} E. Witten, Nucl. Phys. B202 (1982) 253. 
\item{26.} D. Amati, K. Konishi, Y. Meurice, G.C. Rossi and 
G. Veneziano, Phys. Rep. 162 (1988) and references therein. 

\end